%% file: arr.tex
\documentclass[11pt]{article} 
\usepackage[final]{acl}

\usepackage{times}
\usepackage{latexsym}

\input{math_commands.tex}

\usepackage[T1]{fontenc}
\usepackage[utf8]{inputenc}
\usepackage{microtype}
\usepackage{inconsolata}
\usepackage{booktabs}
\usepackage{hyperref}
\usepackage{url}
\usepackage{graphicx}
\usepackage{subfiles}
\usepackage[many]{tcolorbox}
\usepackage{multicol}
\usepackage{makecell}
\usepackage{enumitem}
\usepackage{tabularx}
\usepackage{amsmath,amssymb} 

\definecolor{main}{HTML}{8dbbbb}    
\definecolor{sub}{HTML}{d4e9e9}     

\tcbset{
    sharp corners,
    colback = white,
    before skip = 0.2cm,    
    after skip = 0.5cm      
} 

\newtcolorbox{boxJ}{
    sharpish corners, 
    colback = sub, 
    colframe = main, 
    boxrule = 0pt, 
    toprule = 4.5pt, 
    enhanced,
    fuzzy shadow = {0pt}{-2pt}{-0.5pt}{0.5pt}{black!35} 
}

\title{\includegraphics[width=0.03\linewidth]{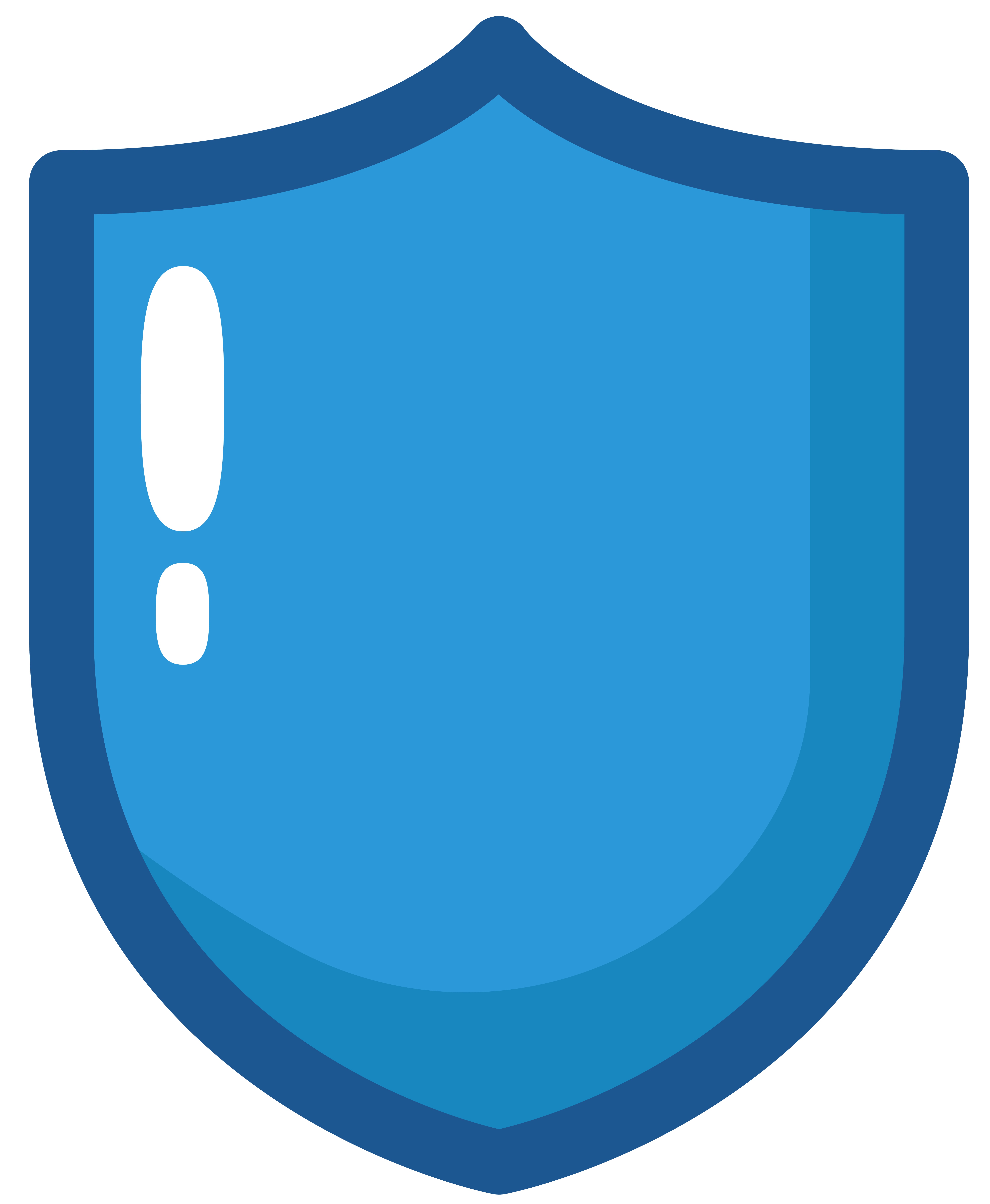} CiteGuard: Faithful Citation Attribution for LLMs via Retrieval-Augmented Validation}

\author{
 \textbf{Yee Man Choi\textsuperscript{1}},
 \textbf{Xuehang Guo\textsuperscript{2}},
 \textbf{Yi R. (May) Fung\textsuperscript{3}},
 \textbf{Qingyun Wang\textsuperscript{2}},
 \\
 \textsuperscript{1}University of Waterloo,
 \textsuperscript{2}College of William and Mary,\\
 \textsuperscript{3}University of Illinois Urbana-Champaign,\\
 \textsuperscript{1}ymchoi@uwaterloo.ca \quad \textsuperscript{2}\{xguo15,qwang16\}@wm.edu \quad
 \textsuperscript{3} yifung2@illinois.edu
}

\usepackage{xcolor} 
\NewDocumentCommand{\qingyun}
{ mO{} }{\textcolor{orange}{\textsuperscript{\textit{Qingyun}}\textsf{\textbf{\small[#1]}}}}

\NewDocumentCommand{\xuehang}
{ mO{} }{\textcolor{blue}{\textsuperscript{\textit{Xuehang}}\textsf{\textbf{\small[#1]}}}}

\begin{document}

\maketitle

\begin{abstract}
\subfile{sections/000_abstract}
\end{abstract}

\section{Introduction}

\subfile{sections/010_intro}



\section{CiteGuard}
\subfile{sections/040_citeguard}

\section{Experiments}
\subfile{sections/050_experiments}

\section{Results}
\subfile{sections/055_experiment_results}

\section{Analysis}
\subfile{sections/060_ablations}

\subfile{sections/020_related_work}

\section{Conclusion and Future Directions}
\subfile{sections/070_conclusion}

\section*{Limitations}
\subfile{sections/080_limitations}
\section*{Ethical considerations}
\subfile{sections/999_ethical_considerations}
\subsubsection*{Acknowledgments}
\subfile{sections/acknowledgements}

\bibliography{arr}

\appendix
\subfile{sections/090_appendix}

\end{document}

%% file: math_commands.tex
\usepackage{amsmath,amsfonts,bm}

\def\eqref#1{equation~\ref{#1}}

\def\1{\bm{1}}

\DeclareMathAlphabet{\mathsfit}{\encodingdefault}{\sfdefault}{m}{sl}
\SetMathAlphabet{\mathsfit}{bold}{\encodingdefault}{\sfdefault}{bx}{n}



%% file: sections/000_abstract.tex
Large Language Models (LLMs) have emerged as powerful assistants for scientific writing. However, concerns remain about the quality and reliability of the generated text, including citation accuracy and faithfulness. While most recent work relies on methods such as LLM-as-a-Judge, the reliability of LLM-as-a-Judge alone is also in doubt. In this work, we reframe citation evaluation as a problem of citation attribution alignment, which assesses whether LLM-generated citations match those a human author would include for the same text. We propose \textit{CiteGuard}, a retrieval-aware agent framework designed to provide more faithful grounding for citation validation. \textit{CiteGuard} improves over the prior baseline by 10 percentage points and achieves up to 68.1\% accuracy on the CiteME benchmark, approaching human performance (69.2\%). It also identifies alternative valid citations and demonstrates generalization ability for cross-domain citation attribution.\footnote{Our code is available at \url{https://github.com/KathCYM/CiteGuard}.}

%% file: sections/010_intro.tex
\noindent
``\textit{If I have seen further than others, it is by standing upon the shoulders of giants}'' --- Isaac Newton. \\
\begin{figure}[h]
    \centering

    \begin{minipage}{\linewidth}
        \centering
        \includegraphics[width=\textwidth]{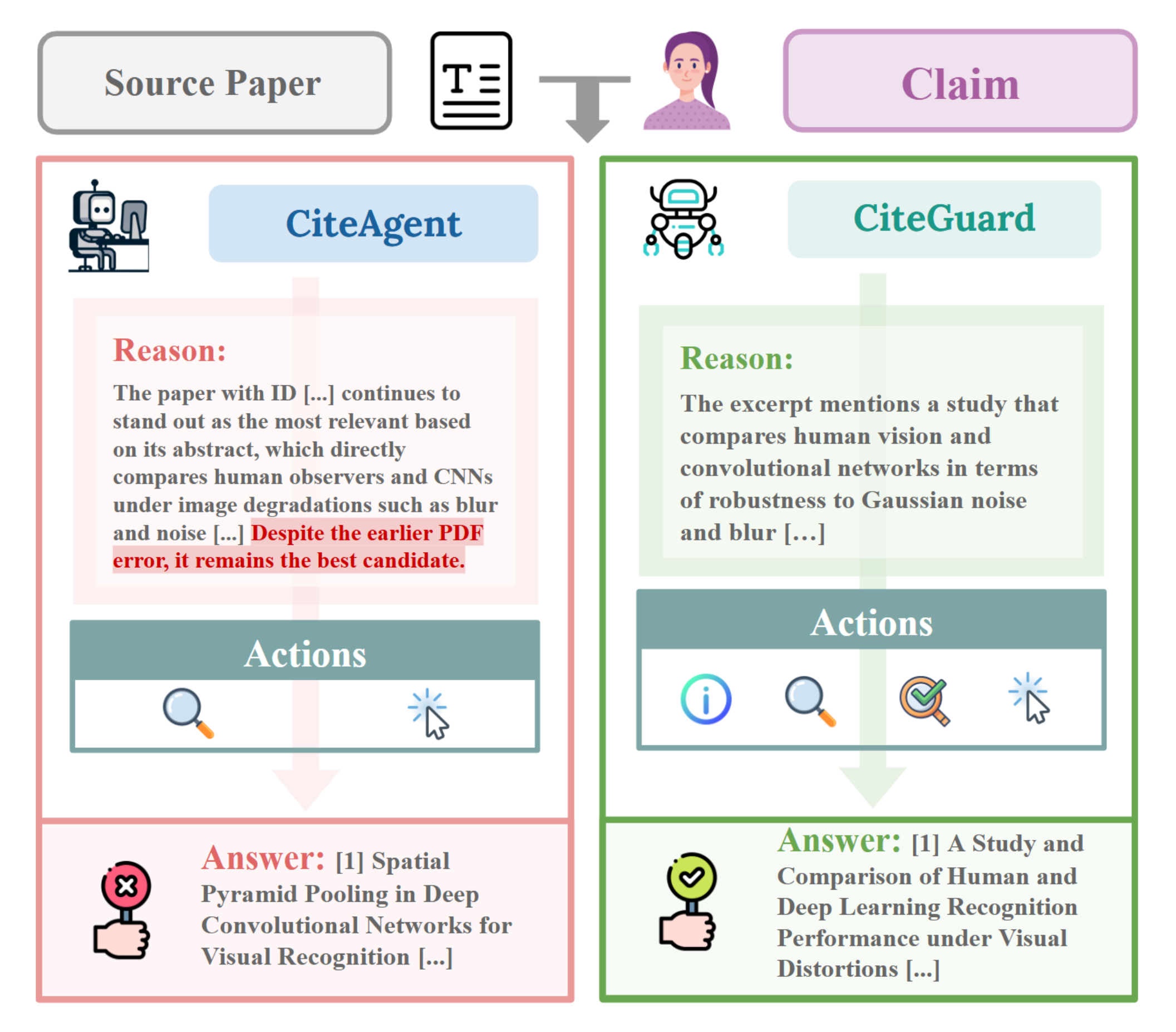}
        \caption{%
        \textit{CiteGuard} succeeds through expanded retrieval actions, whereas CiteAgent~\citep{press2024citeme} fails due to OpenPDF access error.
        }
        \label{fig:citeagent_failure}
\end{minipage}
\end{figure}

\begin{figure*}[t]
\vspace{-10pt}
\begin{center}
\includegraphics[width=\linewidth]{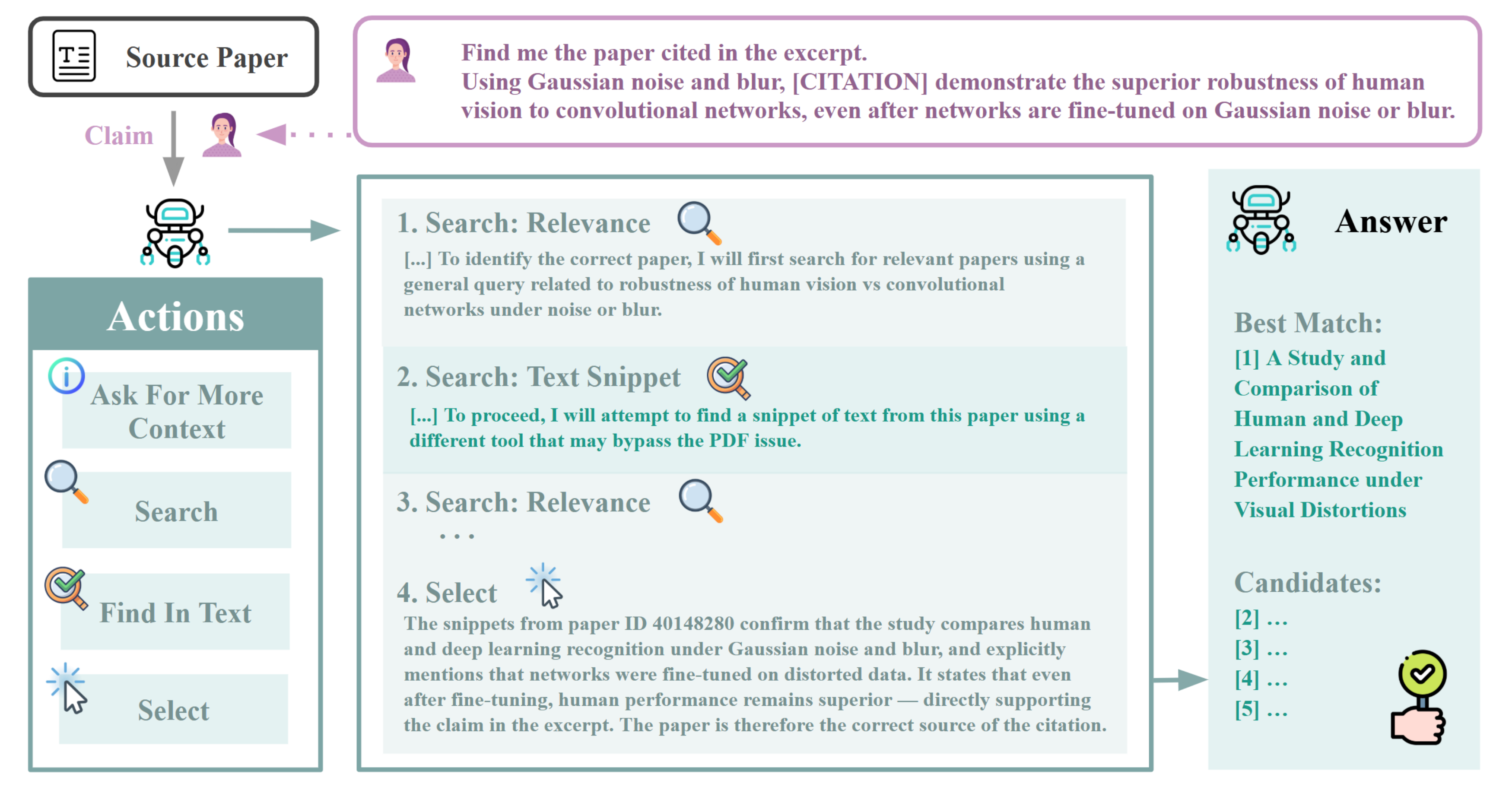}
\end{center}
\vspace{-12pt}
\caption{\textit{CiteGuard} takes an excerpt and performs searches to find a paper that best matches the missing citation. 
}
\vspace{-10pt}
\end{figure*}
Scientific research often progresses by building on the foundation of prior knowledge. Therefore, a thorough and faithful literature review and citation attribution of claims are essential to understand the history and scope of a subject area, and ensure that new findings are properly contextualized~\citep{salton1979citation, snyder2019literature, chigbu2023science}. However, conducting such practices has been increasingly difficult due to the rapid growth in the number of scientific publications~\citep{larsen2010rate, bornmann2015growth}. Recently, over 50 citation hallucinations were found in 300 ICLR 2026 submissions~\citep{Shmatko_Cui_Adam_Esau_2025}. 

Large Language Models (LLMs) and LLM agents have emerged as potentially useful tools to alleviate the burden of researchers and support scientific writing~\cite{lu2024ai, yamada2025ai, asai2024openscholar, wang2025scholarcopilot}. One of the main concerns is hallucinations in LLMs~\citep{ji2023survey, huang2025survey}. For instance, LLMs can generate up to 78-90\% fabricated citations~\citep{asai2024openscholar} and misattribute findings to incorrect sources~\citep{walters2023fabrication}. 

\begin{figure}[ht]
    \centering
    \vspace{0.5em}
    \begin{minipage}{\linewidth}
        \centering
        \setlength{\tabcolsep}{9pt}
        \small
        \begin{tabular}{l|c|c|c}
        \toprule
        \textbf{Method} & \textbf{Precision} & \textbf{Recall} & \textbf{F1} \\
        \midrule
        Zero-shot abstract   & 1.0 & 0.17 & 0.29 \\
        Few-shot abstract    & 1.0 & 0.16 & 0.27 \\
        Zero-shot full text  & 1.0 & 0.36 & 0.53 \\
        Few-shot full text   & 1.0 & 0.38 & 0.55 \\
        \bottomrule
        \end{tabular}
        \captionof{table}{ChatGPT-4o accuracy on citation attribution in the CiteME benchmark.}
        \label{citeme-citation-attribution}
    \end{minipage}
    \vspace{-10pt}
\end{figure}

Retrieval-augmented generation~\citep{lewis2020retrieval,fan2025endtoend} has been proposed to mitigate hallucinations in LLM by retrieving external knowledge to validate the generated text during training data preparation or at inference time~\citep{wang2024autosurvey, asai2024openscholar, wang-etal-2024-scimon, wang2025scholarcopilot}. LLM-as-a-Judge is often used to prepare training data~\citep{asai2024openscholar,asai2024self} or to evaluate generated text~\citep{asai2024openscholar, wang2024autosurvey,he2025verir1precisefaithfulclaim} as it is more scalable in practice, despite the risk of bias and overdependence on LLMs' capabilities~\citep{ye2024justice, thakur2024judging}. LLM-as-a-Judge often assumes that the 
retrieved knowledge used for the generation is available, limiting the use case to evaluating retrieval-augmented output. Moreover, it does not account for situations where the evaluation requires grounding~\citep{krumdick2025no}, such as broader textual understanding, cross-referencing multiple sources, or interpreting ambiguous claims.

We conduct an evaluation of the reliability of LLM-as-a-Judge for citation attribution of human-written scientific claims and their references. 
Although LLMs can recognize apparently incorrect citations, they often reject correct citations due to limited domain-specific knowledge, resulting in a recall as low as 16-17\% (Table~\ref {citeme-citation-attribution}). For instance, LLMs are sensitive to minor variations in terminology (example in App.~\ref{llm_as_a_judge_failures}). This could potentially lead to incorrect evaluation of existing methods and limit the performance of trained LLMs when the training data are filtered using LLM-as-a-Judge. 

We propose \textit{CiteGuard}, an agent that provides more faithful and generalizable citation attribution through retrieval-augmented validation. Prior work, such as CiteAgent~\citep{press2024citeme}, aims to identify citations for scientific claims and improves over direct prompting, but its accuracy still falls short of human performance. \textit{CiteGuard} extends this line of work with two key capabilities: retrieving additional context for the scientific claim and searching directly within paper contents for stronger evidence. Under the same GPT-4o setting, \textit{CiteGuard} improves accuracy by 10 percentage points over CiteAgent. When paired with DeepSeek-R1~\citep{guo2025deepseek}, \textit{CiteGuard} achieves 68.1\% accuracy, approaching human performance at 69.2\%. Human evaluation further shows that \textit{CiteGuard} can identify relevant alternative citations beyond the original benchmark references. Exploratory experiments also suggest that these abilities can generalize to cross-domain and long-paragraph settings. Our contributions are:

\begin{itemize}[itemsep=0pt, parsep=0pt, topsep=0pt, partopsep=0pt]
    \item We propose \textit{CiteGuard}, an agent that provides faithful citation attribution by suggesting multiple appropriate references.
    \item We conduct a detailed analysis and collect human annotations of alternative citations that are not captured by the current CiteME benchmark. We also collect 50 new samples to extend the benchmark to cover cross-domain (biomedical, physics, and math) and long paragraph scenarios, denoted as \textbf{CiteMulti}.
    \item We conduct experiments to show that \textit{CiteGuard} significantly improves accuracy in finding the correct reference, suggesting relevant alternative citations, and that these abilities have the potential to be generalized to cross-domain and long paragraph scenarios.
\end{itemize}

%% file: sections/040_citeguard.tex

\subsection{Problem Formulation}

We formulate the task of finding reference(s) for N excerpts ${x_1,x_2,...,x_N}$ given a pool of $n$ possible reference candidates ${r_1, r_2, ...r_n}$. We have a ground-truth labeling function $y(x_i)$ that can map any excerpt $x_i$ to a ground-truth reference $r^*: y(x_i) = r^*$. We also have another labeling function $\hat{y}(x_i)$ from human annotations that can map any excerpt $x_i$ to a set of k ground truth references $\hat{r}^* = {\hat{r_1}^*,...,\hat{r}_k^*}$: $\hat{y}(x_i) = \hat{r}^*$. This is different from the CiteME~\citep{press2024citeme} setting, where there is only one ground truth reference. \\ The goal of \textit{CiteGuard} is to find a mapping function $f_\theta$ such that 
$f_\theta(x_i) \approx y(x_i), \forall i = {1,...,N}$.\\
The accuracy is defined as:
\begin{align}
    Acc(f_\theta) = \frac{1}{N}\sum_{i=1}^N 1[f_\theta(x_i) = y(x_i)]
\end{align} 
The agreement is defined as:
\begin{align}
Agree(f_\theta) = \frac{1}{N}\sum_{i=1}^N 1[f_\theta(x_i) \cap \hat{y}(x_i) \neq \emptyset]
\end{align}

\subsection{Reference Retrieval}
\label{reference_retrieval}
\begin{figure*}[h]
    \centering
    \includegraphics[width=\textwidth]{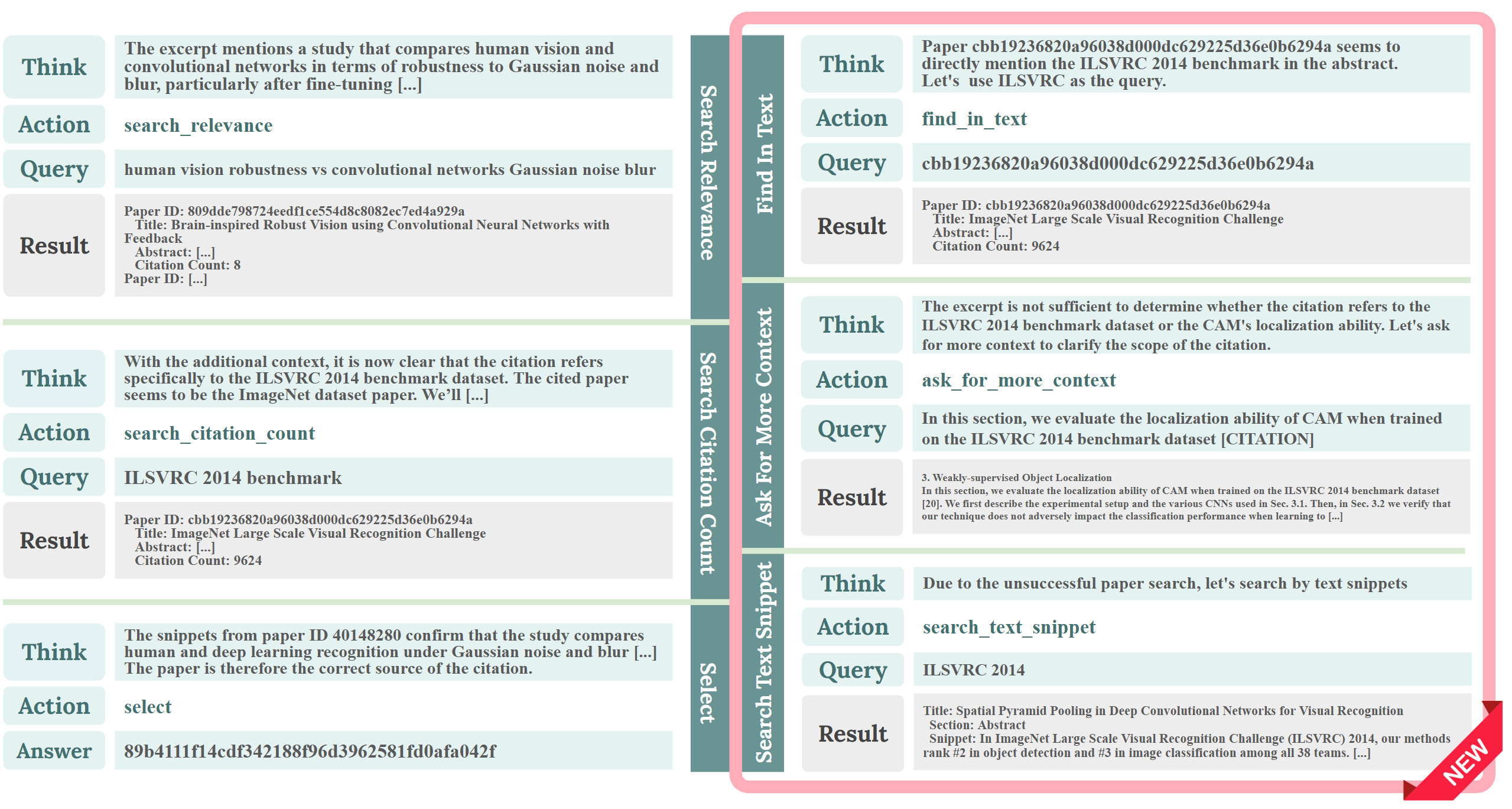}
    \caption{Retrieving Actions. We define six retrieval actions to ensure the efficiency and accuracy of CiteGuard.}
    \label{fig:retrieving_actions}
\end{figure*}

To obtain $f_\theta$, \textit{CiteGuard} introduces new actions in addition to CiteAgent~\citep{press2024citeme}'s actions list (\texttt{search\_citation\_count, search\_relevance, select}). We provide the set of actions below (examples and prompts used can be found in App.~\ref{app:citeguard}). These actions are executed in a paper database $D$ (i.e., Semantic Scholar), which we can query using a search query $q$, and the search result will be appended to $R$. Each paper $P \in D$ contains a title and abstract content $t \in P$, and a body content, with text snippets denoted as $p_i \in P, \forall i$. The source paper that contains the excerpt is $S$. We present some of the examples for the actions in Figure~\ref{fig:retrieving_actions}.  

\noindent1. \textbf{(search\_)citation\_count/relevance} (adopted):\\ Search for a query in the title and abstract fields, then sort the results by citation count/relevance, defined as 
\begin{align*}
    \text{Search}_{c}(q, D) = \text{argsort}_{P \in D}(count(t)) \\
\text{Search}_{r}(q, D) = \text{argsort}_{P\in D}(rel(q,t))
\end{align*}
\noindent2. \textbf{select} (adopted): Select a paper from the search results, defined as 
\begin{align*}
    \text{Select}(P\in R)
\end{align*}
\noindent3. \textbf{find\_in\_text}: Search for a query string within the full text of a specified paper, defined as 
\begin{align*}
    \text{Search}_{t}(q,P) = \text{argsort}_{p\in P}(rel(q,p))
\end{align*}\\
\noindent4. \textbf{ask\_for\_more\_context}: Retrieve the context for an excerpt from the source paper, defined as 
\begin{align*}
    \text{Search}_{cont}(q_i,S) = \{q_{i-3},...,q_{i+3}\}, q_i\in S
\end{align*}\\
\noindent5. \textbf{search\_text\_snippet}: Search for a query string in the full text of papers, defined as 
\begin{align*}
    \text{Search}_{sni}(q,D) = \text{argsort}_{p\in P, P\in D}(rel(q,p))
\end{align*}

\subsection{Iterative Retrieval}
Apart from finding only one reference, \textit{CiteGuard} can suggest multiple references when appropriate to provide a better understanding of the current literature and facilitate comparative analysis. Every run of \textit{CiteGuard} suggests one appropriate reference, with subsequent runs searching for a new appropriate reference. A researcher using this agent can manually audit this iterative process and decide when to stop or allow the agent to make the decision (i.e., CiteGuard would refuse to suggest a paper after exceeding the number of actions, as the papers from the filtered search result are not relevant).

Let $A_k=\{P^{(1)},\ldots,P^{(k)}\}$ denote the set of papers selected during $k$ iterations, and we define the exclusion set $E_k:=A_k$. At iteration $k$, the searchable database is
$D_k:= D \setminus E_{k-1}$, i.e., all retrieval actions that search over papers operate on $D_k$. Concretely, we use the filtered versions of the search actions:
\begin{align*}
\text{Search}^{\neg E}_{c}(q, D) &= \text{argsort}_{P \in D \setminus E}\big(count(t)\big),\\
\text{Search}^{\neg E}_{r}(q, D) &= \text{argsort}_{P \in D \setminus E}\big(rel(q,t)\big),\\
\text{Search}^{\neg E}_{sni}(q, D) &= \text{argsort}_{p\in P,\; P\in D\setminus E}\big(rel(q,p)\big).
\end{align*}
Thus, once a paper $P^{(k)}$ is selected, it is excluded from all subsequent search\_citation\_count,
search\_relevance, and search\_text\_snippet actions by updating $E_k \leftarrow E_{k-1}\cup\{P^{(k)}\}$.


%% file: sections/050_experiments.tex
\subsection{Benchmarking Citation Attribution for Computer Science Excerpts}
We evaluate \textit{CiteGuard} on CiteME~\citep{press2024citeme}, which contains 130 excerpts collected from human-written manuscripts in different Computer Science domains (i.e., computer vision, natural language processing, algorithms, theory), where each excerpt contains exactly one missing citation. The task is for the LLM agent to suggest an appropriate paper to fill in the missing citation. 

\begin{table*}[h]
\vspace{-5pt}
\centering
\small
\setlength{\tabcolsep}{4pt} 
\begin{tabular}{
    l|
    >{\centering\arraybackslash}p{2.2cm}
    >{\centering\arraybackslash}p{2.2cm}
    >{\centering\arraybackslash}p{2.2cm}
    >{\centering\arraybackslash}p{2.1cm}|
    >{\centering\arraybackslash}p{2.3cm}
}
\toprule
& \textbf{Easy(\%)} & \textbf{Medium(\%)} & \textbf{Med-Hard(\%)} & \textbf{Hard(\%)} & 
\textbf{All(\%) ± $\sigma$} \\
\midrule
CiteAgent+GPT-4o  & 69.7 &	52.2 &	15.4 &	2.9 & 35.4 ± 2.3\%\\
\midrule
CiteGuard+GPT-4o & 86.4 & 65.2 & 20.5 & 6.5& 45.1 ± 2.7\%\\
CiteGuard+DeepSeek-R1 & \textbf{95.5} & \textbf{87.0} & \textbf{71.8} & \textbf{15.2} & \textbf{68.1 ± 5.8\%
} \\
CiteGuard+Gemini & 81.8 & 43.5 & 16.7 & 0.0& 34.2  ± 2.7\%\\
CiteGuard+Kimi-K2 & 90.9 & 83.3 & 41.9 & 13.0 & 59.7 ± 2.1\% \\
CiteGuard+Qwen3 & 81.6 & 68.8 & 43.6 & 10.1 & 53.1 ± 3.5\%  \\
\midrule
Human & 95.5& 80.4&69.2&21.7&69.2 \\
\bottomrule
\end{tabular}
\vspace{-3pt}
\caption{CiteGuard accuracy in the CiteME benchmark averaged over 5 runs.}
\label{tab:citeguard}
\vspace{-10pt}
\end{table*}

\subsection{Benchmarking Citation Attribution for Cross-Domain Excerpts}
To explore \textit{CiteGuard}'s robustness on other scenarios, we collect 10 long multi-citation paragraphs from Computer Science literature in addition to the CiteME excerpts. For cross-domain robustness, we collect 20 additional excerpts from the biomedical domain, each containing a single missing citation. The test samples are collected from PubMed and manually verified for availability on Semantic Scholar to ensure solution feasibility. Test samples cover various biomedicine domains, including cancer, epidemiology, microbiome, and microbial genomics. We then collect 20 additional excerpts from the physics and math domains from ArXiv. The combined dataset is denoted as \textbf{CiteMulti}.

\subsection{Evaluation Strategies}
For all the evaluations, we follow the same hyperparameter settings (e.g., temperature) as CiteAgent~\citep{press2024citeme}.

We evaluate \textit{CiteGuard} on both closed- and open-source models, including non-reasoning (GPT-4o~\citep{hurst2024gpt}, Kimi-K2~\citep{team2025kimi}, Qwen3~\citep{yang2025qwen3}, Gemini 2.0 Flash~\citep{Gemini}) and reasoning model (DeepSeek-R1~\citep{guo2025deepseek}), taking the average of 5 run results. 

\begin{table}[thb]
\centering
\small
\setlength{\tabcolsep}{4pt} 
\begin{tabular}{
    l|
    >{\centering\arraybackslash}p{1.5cm}
    >{\centering\arraybackslash}p{1.6cm}
    >{\centering\arraybackslash}p{2.0cm}
}
\toprule
Comparison & $\Delta$ Acc (\%) & $P(\Delta > 0)$ & 95\% CI \\
\midrule
vs GPT-4o & +29.2	& \textbf{1.00}	& [19.0, 37.8] \\
vs Gemini & +35.4	& \textbf{1.00}	& [24.3, 44.2] \\
vs Qwen3  & +31.5	& \textbf{1.00}	& [21.6, 39.8] \\
vs Kimi-K2 & +19.2	& \textbf{0.99}	& [9.4, 28.0] \\
\bottomrule
\end{tabular}
\vspace{-5pt}
\caption{Bayesian paired comparisons (vs DeepSeek-R1) among the CiteGuard variants.}
\label{tab:bayes}
\vspace{-10pt}
\end{table}

\begin{table}[htb]
\centering
\small
\setlength{\tabcolsep}{3pt} 

\begin{tabular}{
    l|
    >{\centering\arraybackslash}p{2.9cm}
    >{\centering\arraybackslash}p{2.9cm}
}
\toprule
Method & \textbf{BioMed(\%)} & \textbf{Long Para.(\%)} \\
\midrule
CiteAgent & 26.6 & 40.0 \\
CiteGuard & 28.3 & 46.6 \\
\midrule
Method & \textbf{Phy\&Math(\%)} & \textbf{All(\%)} \\
\midrule
CiteAgent & 52.5 & 39.6 \\
CiteGuard & 55.0 & 42.6 \\
\bottomrule
\end{tabular}
\vspace{-5pt}
\caption{Agent accuracy in the CiteMulti benchmark, averaged over 3 runs using Kimi-K2.}
\vspace{-20pt}
\label{tab:citemulti}
\end{table}

\subsection{Difficulty Level Labels}

We label the sample with difficulty levels using the following criteria from the results of a single run:
\begin{itemize}[itemsep=3pt, parsep=0pt, topsep=3pt, partopsep=0pt]
    \item Easy (22 excerpts): Correct for all models
    \item Medium (46 excerpts): Correct for more than three out of five models
    \item Medium-Hard (39 excerpts): Correct for no more than two out of five models
    \item Hard (23 excerpts): Incorrect for all models
\end{itemize}
We present some examples of excerpts in the more difficult levels in App.~\ref{app:difficulty_levels}.

\subsection{Human Assessment}
To evaluate the quality of alternative citations suggested by CiteGuard, we manually inspect the suggested citations from a single run for all the different settings (see Fig.~\ref{alternative_citation} and Fig.~\ref{alternative_citation2} for examples).
For each sampled claim, we ask at least two expert annotators with backgrounds in computer science and scientific writing to independently judge whether the list of suggested citations produced by CiteGuard, backed by different LLMs, is an appropriate alternative. We defined an alternative citation as ``appropriate” if it provides equivalent or stronger evidence for the scientific claim compared to the original reference.
Inter-annotator agreement reports \textbf{72.7\%}, indicating high consistency among human annotators.



%% file: sections/055_experiment_results.tex
\subsection{CiteGuard Accurately Grounds Scientific Claims Through Enhanced Actions}




Results in Table~\ref{tab:citeguard} demonstrate that \textit{CiteGuard} substantially outperforms CiteAgent, improving the accuracy of retrieving the oracle citation by 10\% on CiteME when both are powered by GPT-4o. When backed by open-source models DeepSeek-R1 and Kimi-K2, \textit{CiteGuard} achieves up to 68.1\% accuracy, approaching the 69.2\% human performance.

We also present the Bayesian paired comparison results among CiteGuard variants using per-sample correctness outcomes from a single run in Table~\ref{tab:bayes}. As shown in the results, DeepSeek-R1 demonstrates strong posterior evidence of improved accuracy over GPT-4o, Gemini, and Qwen3 (with $P(\Delta>0)\ge0.99$ and the 95\% credible intervals above zero), while its advantage over Kimi-K2 is smaller. Since per-sample predictions are unavailable for the original CiteAgent-reported statistics, we re-evaluate CiteAgent under identical experimental settings. CiteGuard achieves a 5-run average of 45.1\% ± 2.7\%, compared to 35.4\% ± 2.3\% for CiteAgent (both using GPT-4o), with posterior $\Delta$ Acc (\%) = +8.46\%, $P(\Delta > 0)$ = 0.98, 95\% CI = [0.28\%, 16.32\%].

This improvement is driven by \textit{CiteGuard}’s extended retrieval actions (\S~\ref{reference_retrieval}), which make citation search more flexible and robust.
As illustrated in Fig.~\ref{fig:citeagent_failure}, while CiteAgent relies heavily on the \texttt{read} action that assumes reliable PDF access, \textit{CiteGuard} succeeds through introducing two key new actions:
(1) \texttt{ask\_for\_more\_context} enables the agent to proactively query for additional claim context when the initial snippet is insufficient, and (2) \texttt{search\_text\_snippet} allows searching directly within paper contents.

There are two key advantages of search\_text\_snippet:

\begin{itemize}[itemsep=3pt, parsep=0pt, topsep=3pt, partopsep=0pt]
    \item When relevant information is not explicit in a paper’s title or abstract, search\_text\_snippet can retrieve matches from the main content, whereas earlier search actions rely solely on title and abstract matching
    \item It does not require PDF access, which is often unavailable or difficult to parse in Semantic Scholar. When PDF access fails, agents may rely only on titles and abstracts and make ill-informed selections. In contrast, search\_text\_snippet retrieves text directly from paper content and excludes papers without accessible content, leading to more robust retrieval.
\end{itemize}
%
%

This step-by-step reasoning, together with more advanced actions, enables \textit{CiteGuard} to accurately identify the oracle citation where CiteAgent fails, improving the accuracy and robustness of scientific claim grounding, particularly in real-world citation retrieval with complex long-range contexts.

\subsection{CiteGuard Effectively Suggests Alternative Citations}




\begin{table*}[thb]
\vspace{-5pt}
\centering
\small
\setlength{\tabcolsep}{4pt}
\begin{tabular}{
    l|
    >{\centering\arraybackslash}p{2.8cm}
    >{\centering\arraybackslash}p{2.8cm}|
    >{\centering\arraybackslash}p{2.8cm}
    >{\centering\arraybackslash}p{2.8cm}
}
\toprule
& \multicolumn{2}{c|}{\textbf{Oracle Coverage}} 
& \multicolumn{2}{c}{\textbf{Agreement}} \\
\cmidrule(lr){2-3} \cmidrule(lr){4-5}
Method 
& \textbf{Oracle-in-Set@1} 
& \textbf{Oracle-in-Set@5} 
& \textbf{Agreement@1} 
& \textbf{Agreement@5} \\
\midrule
CiteAgent+GPT-4o        & 35.4\% & 49.2\% & 50.5\% & 70.1\% \\
CiteGuard+GPT-4o        & 45.1\% & 56.7\% & 62.9\% & 75.3\% \\
CiteGuard+DeepSeek-R1   & \textbf{68.1\%} & \textbf{76.9\%} & \textbf{71.6\%} & 82.6\% \\
CiteGuard+Gemini        & 34.2\% & 42.4\% & 48.4\% & 60.8\% \\
CiteGuard+Kimi-K2       & 59.7\% & 70.8\% & \textbf{71.6\%} & \textbf{87.8\%} \\
CiteGuard+Qwen3         & 53.1\% & 69.6\% & 66.0\% & 85.8\% \\
\bottomrule
\end{tabular}
\vspace{-3pt}
\caption{Oracle coverage and agreement metrics across models. Oracle-in-Set@k measures whether the ground-truth citation appears in the top-$k$ retrieved set, while Agreement@k measures alignment with human-annotated relevant citations.}
\label{tab:oracle_agreement}
\vspace{-10pt}
\end{table*}

Through manual assessment, \textit{CiteGuard} showcases its ability to generate high-quality alternative citations beyond the original reference. Using aggregated human annotations as a new oracle, Table~\ref{tab:oracle_agreement} computes the agreement between \textit{CiteGuard}’s suggested citations and human judgments. Across models, \textit{CiteGuard} achieved substantial alignment with human evaluations, demonstrating its potential to identify relevant alternative literature. 

Notably, this ability is \textbf{model-agnostic}: both proprietary models like GPT-4o and open-source models like Qwen3 can effectively identify relevant alternatives. 
Fig.~\ref{alternative_citation} demonstrates \textit{CiteGuard}'s \textit{backward reasoning} ability based on the excerpt.
Fig.~\ref{alternative_citation2} further shows the \textit{lateral reasoning} capacity of \textit{CiteGuard}, where \textit{CiteGuard} effectively identifies highly related work as the oracle reference suggested. We include some examples and more details about the human assessment process in App.~\ref{human_assessment_on_alternative_citations}.

\subsection{CiteGuard Can Generalize to Cross-Domain Citation Attribution}

Results in Table~\ref{tab:citemulti} demonstrate \textit{CiteGuard}'s potential to generalize its ability to ground scientific claims to cross-domain and long paragraph scenarios. We note that 
the performance gain of CiteGuard over the CiteAgent baseline on CiteMulti is smaller than that observed on CiteME. This is likely due to domain-specific distribution shifts and differences in citation styles. We discuss the limitations associated with the search engine in the limitations section.

%% file: sections/060_ablations.tex

\subsection{Effectiveness of Multi-Run}
\begin{table}[h!]
\centering
\small
\setlength{\tabcolsep}{3pt}
\renewcommand{\arraystretch}{1.1}

\begin{tabular}{
    l|
    >{\centering\arraybackslash}p{5.3cm}
}
\toprule
\textbf{\# of Iterations} & \textbf{Accuracy (\%)} \\
\midrule
1 & 60.0 \\
2 & 70.0\% (+ 10.0\%)\\
3 & 75.4\% (+ 5.4\%)
 \\
\bottomrule
\end{tabular}
\vspace{-5pt}
\caption{\textit{CiteGuard}+Kimi-K2 accuracy difference on the CiteME benchmark when using iterative runs.}
\label{tab:multi-run}
\vspace{-5pt}
\end{table}
To evaluate \textit{CiteGuard}'s performance with iterative run ability enabled, we present the result of CiteGuard + Kimi-K2 on the CiteME benchmark in Table~\ref{tab:multi-run}. The result demonstrates that CiteGuard with iterative runs consistently improves accuracy, with diminishing yet still meaningful gains as the number of iterations increases. For the excerpt where only one particular citation is appropriate and is selected in prior runs, \textit{CiteGuard} sometimes refuses to suggest a paper after exceeding the number of actions, as the papers from the filtered search result are not relevant.

\subsection{Effectiveness of Each Action}
To demonstrate the contribution of each newly added action, we perform an ablation study for the two actions \textbf{ask\_for\_more\_context} and \textbf{search\_text\_snippet} and present the result below. As shown in Table~\ref{tab:action-ablation}, introducing either action leads to a substantial improvement over the CiteAgent baseline, indicating that both additional context elicitation and external evidence retrieval are effective for citation verification. In particular, the \textbf{search\_text\_snippet} action yields a larger individual gain, suggesting that access to retrieved textual evidence is particularly important. When both actions are enabled, CiteGuard achieves the highest accuracy, indicating that the two actions complement each other.

\begin{table}[h!]
\centering
\small
\setlength{\tabcolsep}{3pt}
\renewcommand{\arraystretch}{1.1}

\begin{tabular}{
    l|
    >{\centering\arraybackslash}p{2.5cm}
}
\toprule
\textbf{Method} & \textbf{Accuracy (\%)} \\
\midrule
CiteAgent &	36.2\% \\
CiteGuard (+ ask\_for\_more\_context) & 53.1\% \\
CiteGuard (+ search\_text\_snippet) &	57.7\% \\
CiteGuard (+ both) & 59.7\% \\
\bottomrule
\end{tabular}
\caption{Accuracy difference on the CiteME benchmark when using different sets of actions.}
\label{tab:action-ablation}
\vspace{-12pt}
\end{table}

\subsection{Retrieval vs Long-Context}
To demonstrate the effect of retrieving only relevant parts of the paper versus providing the full paper text, we run the \textit{CiteGuard}+Kimi-K2 agent, replacing the "find\_in\_text" action with the "read" action and present the results in Table~\ref{kimi-read}. With the "read" action, the accuracy increased by 3.07\%, at the cost of 2$\times$ more tokens. The number of tokens can be as large as 4$\times$ as shown in App.\ref{app:long_context}. 

Although reading the full paper content in context can provide some benefits, it is at the cost of significantly more tokens. When using \textit{CiteGuard}, users would need to determine whether to use retrieval or long-context based on the token budget.

\begin{table}[h!]
\centering
\small
\setlength{\tabcolsep}{3pt}
\renewcommand{\arraystretch}{1.1}

\begin{tabular}{
    l|
    >{\centering\arraybackslash}p{2.5cm}|
    >{\centering\arraybackslash}p{3.0cm}
}
\toprule
\textbf{Method} & \textbf{Accuracy (\%)} & \textbf{Avg \# of Tokens} \\
\midrule
read & 63.1 & 33{,}544.68 \\
find\_in\_text & 60.0 & 15{,}451.43 \\
\bottomrule
\end{tabular}
\vspace{-5pt}
\caption{\textit{CiteGuard}+Kimi-K2 accuracy difference on the CiteME benchmark when using different actions to get information from the paper content.}
\label{kimi-read}
\vspace{-5pt}
\end{table}

\subsection{Tradeoffs of Human in the Loop}
\vspace{-5pt}
\begin{table}[h!]
\centering
\small
\setlength{\tabcolsep}{3pt}
\renewcommand{\arraystretch}{1.1}

\begin{tabular}{
    l|
    >{\centering\arraybackslash}p{2.2cm}
}
\toprule
\textbf{Method} & \textbf{Accuracy (\%)} \\
\midrule
CiteGuard  (w/o ask\_for\_more\_context) &	57.7\% \\
CiteGuard  (w/ ask\_for\_more\_context) &	59.7\%
 \\
\bottomrule
\end{tabular}
\vspace{-5pt}
\caption{\textit{CiteGuard}+Kimi-K2 accuracy difference on the CiteME benchmark when using different actions.}
\label{tab:ablation_tradeoff}
\vspace{-5pt}
\end{table}

When we perform the evaluation, ask\_for\_more\_context does not take input from users, but instead returns the paragraph in the source paper that contains the excerpt (with the citation masked). However, for applications where the excerpt is not taken from a paper but provided by a researcher in the process of writing, the user would have to provide some additional context manually if needed.

We perform an ablation study by removing ask\_for\_more\_context action and present the results in Table~\ref{tab:ablation_tradeoff}; the performance difference is subtle (2\%), but may be important, depending on the use case. Although ask\_for\_more\_context and the full list of actions are provided as default, there is an option to remove some of the actions, making CiteGuard applicable to different use cases. 
\subsection{Reasoning vs Non-Reasoning Models}
Table~\ref{tab:citeguard} shows the difference of open-sourced reasoning (DeepSeek-R1) and non-reasoning model (Kimi-K2) in overall performance. 

As demonstrated in the example (Fig.~\ref{ablation_reasoning}), a reasoning model tends to question itself (\texttt{"But note:...However,..."}) and consider other available actions, while a non-reasoning model would be more confident in its action (\texttt{"I can still be confident that..."}). Although the agent backed by both models eventually arrived at different citations, both are deemed correct through human assessment, demonstrating that \textit{CiteGuard} is not dependent on reasoning ability.

\begin{figure}[h]
\begin{boxJ}

DeepSeek-R1: ... But note: the excerpt says "the raw image could be fed directly into the reinforcement learning algorithm through a series of convolutions". This is exactly what DQN did. However, let's break it down:
- The citation is likely for the method of using convolutions to process raw images in RL, not necessarily the entire RL algorithm...
\ \\

Kimi-K2: ...I can still be confident that its abstract already declares it is the first deep learning model to successfully learn control policies directly from high-dimensional sensory input, where input is raw pixels...
\
\end{boxJ}
\vspace{-5pt}
\caption{CiteGuard thought example when backed by reasoning model (DeepSeek-R1) and non-reasoning model (Kimi-K2)}
\label{ablation_reasoning}
\vspace{-25pt}
\end{figure}
\subsection{CiteGuard vs Paper Finders}

\begin{table}[h!]
\centering
\small
\setlength{\tabcolsep}{3pt} 
\renewcommand{\arraystretch}{1.1}

\begin{tabular}{
    l|
    >{\centering\arraybackslash}p{2.2cm}
    >{\centering\arraybackslash}p{2.2cm}
}
\toprule
& \textbf{Top 1} & \textbf{Top 10}\\
\midrule
AI2 Paper Finder & 38.5& 60.0\\
\midrule
Ours+Gemini & 36.9 &  46.2 \\
Ours+DeepSeek-R1 & \textbf{65.4} & \textbf{84.6} \\
\bottomrule
\end{tabular}
\caption{AI2 Paper Finder~\citep{Ai2}'s accuracy (\%) on CiteME compared to CiteGuard.}
\label{paper-finder}
\vspace{-12pt}
\end{table}

\begin{table*}[h]
\centering
\small
\setlength{\tabcolsep}{3pt} 
\renewcommand{\arraystretch}{1.1}

\begin{tabular}{
    l|
    >{\centering\arraybackslash}p{2.8cm}
    >{\centering\arraybackslash}p{2.8cm}
    >{\centering\arraybackslash}p{3.0cm}
    >{\centering\arraybackslash}p{2.8cm}
}
\toprule
Model & Avg. Input Tokens & Avg. Output Tokens & Avg. Cost / Sample (\$) & Platform \\
\midrule
GPT-4o & 17{,}931.8 & 1{,}705.8 & 0.12 & OpenAI \\
DeepSeek-R1 (671B / 37B) & \textbf{15{,}004.9} & 1{,}771.4 & 0.005 & DeepSeek \\
Gemini-2.0-Flash & 19{,}064.9 & 1{,}449.4 & \textbf{0.00} & Google (free tier) \\
Kimi-K2 (1T / 30B) & 15{,}451.4 & \textbf{826.7} & 0.017 & Together AI \\
Qwen3 (235B / 22B) & 14{,}598.8 & 936.8 & \textbf{0.003} & Together AI \\
\bottomrule
\end{tabular}
\caption{Average token usage and API cost per sample for each evaluated model. Numbers are averaged over the evaluation set. Model sizes are reported as total parameters / activated parameters where applicable.}
\label{tab:cost}
\vspace{-5pt}
\end{table*}

\begin{table*}[thb]
\vspace{-5pt}
\centering
\small
\setlength{\tabcolsep}{4pt}
\renewcommand{\arraystretch}{1.1}
\begin{tabular}{
    l|
    >{\centering\arraybackslash}p{2.1cm}
    >{\centering\arraybackslash}p{2.1cm}
    >{\centering\arraybackslash}p{2.2cm}
    >{\centering\arraybackslash}p{2.1cm}|
    >{\centering\arraybackslash}p{2cm}
}
\toprule
& \textbf{Easy(\%)} & \textbf{Medium(\%)} & \textbf{Med-Hard(\%)} & \textbf{Hard(\%)} & \textbf{All(\%)} \\
\midrule
CiteGuard + Semantic Scholar & \textbf{95.5} & \textbf{87.0} & \textbf{71.8} & 15.2 & \textbf{68.1} \\
CiteGuard + arXiv            & 68.2 & 54.3 & 28.2 & \textbf{21.7} & 43.1 \\
\bottomrule
\end{tabular}
\vspace{-3pt}
\caption{Performance comparison across retrieval backends. Semantic Scholar provides stronger performance overall, while arXiv demonstrates the portability of CiteGuard under constrained retrieval capabilities.}
\label{tab:alt_retrieval}
\vspace{-10pt}
\end{table*}

An alternative to finding potential references using \textit{CiteGuard} is to use a paper finder. We run AI2 Paper Finder~\citep{Ai2} on CiteME and present the results in Table~\ref{paper-finder}. AI2 Paper Finder searches and ranks the documents, which can result in a long list of papers, while CiteGuard operates in a setting that only produces one suggestion at a time. Therefore, we report AI2 Paper Finder's accuracy by taking the top k-ranked documents, and CiteGuard's top k performance from k separate (non-iterative) runs. We argue that \textit{CiteGuard} matches Paper Finder in terms of accuracy, if not surpassing it. In particular, the top 10 accuracy of AI2 Paper Finder is 5.4 percentage points below the top 1 accuracy of CiteGuard+DeepSeek-R1, demonstrating that \textit{CiteGuard} is more reliable, which is likely because it incorporates the context of the excerpt.

\subsection{Cost Analysis}

We provide the average number of tokens and the API cost per sample for each model we evaluated in Table~\ref{tab:cost}. Overall, CiteGuard incurs relatively modest inference cost. The best-performing model, DeepSeek-R1, has an average API cost of approximately \$0.005 per citation, which is substantially lower than GPT-4o (\$0.12 per citation). These results suggest that CiteGuard’s performance improvements are achieved without incurring disproportionate inference cost, supporting its practicality and scalability.

\subsection{Alternative Retrieval Pipeline}

To demonstrate the portability of CiteGuard’s action space, we implement an alternative retrieval backend using ArXiv. Due to API limitations (e.g., lack of citation counts and snippet-based search), we disabled the corresponding actions for this backend. Using DeepSeek-R1 as the LLM backend, the results are shown in Table~\ref{tab:alt_retrieval}.

While performance differs due to backend capabilities and coverage, these results demonstrate that CiteGuard can be adapted to alternative retrieval systems. We view this as an encouraging step toward using CiteGuard as a testbed for designing improved citation attribution-oriented retrieval backends.

%% file: sections/020_related_work.tex
\section{Related Work}
\label{app:related_work}
\subsection{Retrieval-Augmented Generation and LLMs for Scientific Research}
Retrieval Augmented Generation (RAG) models were first introduced as models that can combine parametric and non-parametric memory~\citep{lewis2020retrieval}. Recently, RAG has shown to be a promising direction toward mitigating hallucinations and other challenges in knowledge-intensive tasks for LLMs~\citep{borgeaud2022improving, gao2023retrieval}. One application is the use of LLMs/LLM Agents to assist human researchers, such as knowledge discovery, proposing ideas, carrying out experiments, scientific writing, conducting reviews, or even transforming paper into interactive agents~\citep {lu2024ai, yamada2025ai, miao2025paper2agent}. As part of the effort to mitigate the hallucination issue in LLMs for scientific writing, RAG-aware fine-tuned LLMs for literature summaries have been introduced~\citep{asai2024openscholar, wang2025scholarcopilot}.

\subsection{Citation Suggestion}
There have been different approaches to citation recommendation before the era of LLMs, including information retrieval~\citep{farber2020hybridcite} and neural networks~\citep{ebesu2017neural, yang2018lstm, jeong2020context}. These methods require re-training and do not account for the rapidly updating paper database. In light of this, LLM agentic workflow~\citep{press2024citeme} has been proposed to enable access to a real-time paper database.

In this work, we adopted the CiteAgent~\citep{press2024citeme} framework, where retrieval is performed through tool calls to the Semantic Scholar API, which we treat as a black box for retrieval. This approach would benefit from further improvements in the retrieval pipeline of the API. The framework is built to enable multiple rounds of retrieval and reading, with the choice of action dependent on the agent's own decision following its thought, similar to the ReAct approach~\citep{yao2023react}. 
\textit{CiteGuard} builds on this retrieval-based citation recommendation setting to provide more faithful citation attribution and to surface alternative valid references.

\subsection{LLM-as-a-Judge}
Evaluation of LLM-generated text has traditionally been carried out by humans. Collecting human annotations is costly and not scalable. To overcome this issue, LLM-as-a-Judge was introduced to automate the evaluation process~\citep{zheng2023judging}. Due to the improved scalability, LLM-as-a-Judge has been widely used to evaluate LLM-generated scientific writing. For instance, OpenScholar~\citep{asai2024openscholar} uses LLM-as-a-Judge to filter and refine LLM-synthesized training data. However, LLM-as-a-Judge exhibits bias~\citep{ye2024justice, gu2024survey} or sensitivity to prompt~\cite{thakur2024judging}.  Moreover, LLM-as-a-Judge often requires a text snippet of the citation under review, which limits its use case for scenarios where the text snippets used during generation are not available. In this work, we explore expanding LLM-as-a-Judge to include RAG to alleviate biases and provide a more robust evaluation in cases where relevant text snippets are not directly available. A similar idea is Agent-as-a-Judge~\citep{zhuge2024agent} targeting the task of automated code generation for AI development. 

%% file: sections/070_conclusion.tex
We observe the limitation in using LLM-as-a-Judge for citation attribution of scientific writing and propose \textit{CiteGuard} agent to provide a more faithful citation attribution through retrieval-augmented validation. We show that \textit{CiteGuard} approaches human performance in retrieving correct citations, while its suggested alternative citations are also judged relevant by human annotators. We further demonstrate that \textit{CiteGuard} has promising generalization ability across domains.

Faithful attribution is critical to the integrity of scientific communication, and the research community is placing increasing emphasis on accurate, verifiable citation attribution. We envision \textit{CiteGuard} as one of the components in future scientific writing and review pipelines, to mitigate hallucinated citations and promote more trustworthy scholarly literature.

%% file: sections/080_limitations.tex
Currently, the implementation of CiteGuard is based on the Semantic Scholar API, which causes CiteGuard's performance to be limited by the coverage of the database and the ability of the retrieval pipeline of the API. One future direction of CiteGuard is enabling the use of other research literature databases and retrieval pipelines. Although we have shown that CiteGuard agent works well with both open-sourced and closed-sourced, both reasoning and non-reasoning models, we have not yet explored its performance on smaller open-sourced models (e.g., models with less than 1B parameters) due to the limitation of time. We plan to conduct such an analysis and evaluate how much CiteGuard depends on the models' size.

%% file: sections/999_ethical_considerations.tex
Our work aims to promote a more faithful citation attribution for scientific writing, regardless of machine-generated or human-generated. The framework relies on Large Language Models, which may exhibit systemic biases in research communities, such as geographic and linguistic biases. Although our method is model-agnostic, we acknowledge that mitigating these biases is still an open challenge. Future work includes better representation of under-cited or non-English sources.  Our framework uses Semantic Scholar, which is an open-access research tool for scientific literature, through its API. We have not used any private or sensitive data. All human annotators (including the authors) participated in a voluntary manner, with their identities kept anonymous during the analysis. 

%% file: sections/acknowledgements.tex
We thank Semantic Scholar and Arxiv for providing APIs to query their database, and Google Gemini for API access to Gemini-2.0-Flash. We thank Tom Hope and Freda Shi for valuable discussions and insightful feedback that helped improve this work. We sincerely thank our annotators for their annotations. The authors acknowledge William \& Mary Research Computing for providing computational resources and/or technical support that have contributed to the results reported within this paper.

%% file: sections/090_appendix.tex
\section{CiteGuard}
\label{app:citeguard}
\subsection{Prompts}
The system prompt (Fig.~\ref{citeguard: system_prompt}) and examples provided in the prompt for each newly added actions (Fig. \ref{citeguard: ask_more_context}, Fig. \ref{citeguard: find_in_text} and Fig. \ref{citeguard: search_text_snippet}) are presented below.


\begin{figure*}[htb!]
\begin{boxJ}
You are given an excerpt from a paper, where a citation was deleted. I'm trying to find the citation (ignore the word [CITATION], that's just where the citation was deleted from. You will be asked to help me find the paper from which the citation was deleted. You are equipped with the following tools that will help you in your task: you can search, you can select to find a keyword in a paper from the search results, or you can select a paper as your final answer.

<FORMAT\_INSTRUCTIONS>

Your responses have to include one of the actions above.
Before you take any action, provide your thoughts for doing so.
Do not include anything other than your thoughts and an action in your responses. You must include exactly one action in your responses.

Keep in mind that you can only find keyword in papers or select papers after you search. You can always search, and then search again. You can also find different keywords in a few papers consecutively, without searching again (as long as the papers appeared in your last search). 

If your search does not return any relevant results, please try the following:
1. Adjust your query to focus on individual parts of the claim separately, rather than the entire sentence. It is likely that the citation supports only the immediate preceding concept, not the full claim. Use simpler, more general search queries with fewer words (around 3). 
2. Avoid overly specific or plural terms—use base forms of key concepts to improve match quality. For example, ``We apply contrastive learning to improve the representation quality of a ResNet-based encoder [CITATION].", the citation is most likely for ResNet instead of covering both contrastive learning and ResNet.
3. Change to search by text snippets. You should respond with the following, and use the paper title to do subsequent search.
\end{boxJ}
\caption{CiteGuard System Prompt}
\label{citeguard: system_prompt}
\vspace{-12pt}
\end{figure*}

\begin{figure*}[htb]
\begin{boxJ}
The excerpt is from paper title "Learning Deep Features for Discriminative Localization":
In this section, we evaluate the localization ability of CAM when trained on the ILSVRC 2014 benchmark dataset [CITATION]

You would respond with:
\{
    "reason": "The excerpt is not sufficient to determine whether the citation refers to the ILSVRC 2014 benchmark dataset or the CAM's localization ability. Let's ask for more context to clarify the scope of the citation.",
    "action": \{
        "name": "ask\_for\_more\_context",
	"query": "In this section, we evaluate the localization ability of CAM when trained on the ILSVRC 2014 benchmark dataset [CITATION]"
	"paper\_title": "Learning Deep Features for Discriminative Localization"
    \}
\}

You will then be given the following input:
3. Weakly-supervised Object Localization
In this section, we evaluate the localization ability of CAM when trained on the ILSVRC 2014 benchmark dataset [20]. We first describe the experimental setup and the various CNNs used in Sec. 3.1. Then, in Sec. 3.2 we verify that our technique does not adversely impact the classification performance when learning to localize and provide detailed results on weakly-supervised object localization.

You would respond with:
\{
    "reason": "With the additional context, it is now clear that the citation refers specifically to the ILSVRC 2014 benchmark dataset. The cited paper seems to be the ImageNet dataset paper. We’ll do a search to make sure.",
    "action": \{...\}
\}
\end{boxJ}
\caption{CiteGuard ask\_for\_more\_context Prompt}
\label{citeguard: ask_more_context}
\vspace{-12pt}
\end{figure*}

\begin{figure*}[h!]
\begin{boxJ}
You would then reply:
\{
    "reason": "Paper cbb19236820a96038d000dc629225d36e0b6294a seems to directly mention the ILSVRC 2014 benchmark in the abstract. Though we can use ILSVRC 2014 as the query, but fewer words is better, so we only use ILSVRC as the query.",
    "action": \{ 
    "name": "find\_in\_text",
        "paper\_id": "cbb19236820a96038d000dc629225d36e0b6294a"
        "query": "ILSVRC"
    \}\}
    
You would then be given the following input:\\
Based on this work, we attended the competition of ILSVRC 2014 [26]
\end{boxJ}
\caption{CiteGuard find\_in\_text Prompt}
\label{citeguard: find_in_text}
\vspace{-12pt}
\end{figure*}

\begin{figure*}[h!]
\begin{boxJ}
 You should respond with the following, and use the paper title to do subsequent search:
\{
    "reason": "Due to the unsuccessful paper search, let's search by text snippets instead",
    "action": \{
        "name": "search\_text\_snippet",
        "query": "ILSVRC 2014"
    \}
\}
You will then be given the following input:
- Title: Spatial Pyramid Pooling in Deep Convolutional Networks for Visual Recognition
  Section: Abstract
  Snippet: In ImageNet Large Scale Visual Recognition Challenge (ILSVRC) 2014, our methods rank \#2 in object detection and \#3 in image classification among all 38 teams.
\end{boxJ}
\caption{CiteGuard search\_text\_snippet Prompt}
\label{citeguard: search_text_snippet}
\vspace{-12pt}
\end{figure*}


\section{Examples of CiteGuard Short Trajectories}
\label{app:contamination}
We evaluate the risk of contamination, which means the models are aware of the citation beforehand and do not use search tools to accomplish the task. We manually select some successful short trajectories that are more likely to be an indication of contamination and put the examples in Fig. \ref{citeguard: deepseek_short} and ~\ref{citeguard: kimi_short}. Although these successful trajectories are short, we have not found any instances where the agent knows the ground-truth citation in advance and directly searches for the target citation. Instead, in these trajectories, both agents compose a generic search query and identify the appropriate references from the list of search results. To ensure that CiteGuard’s performance is driven by its retrieval-augmented reasoning rather than internal memorization, we do not simply audit for direct ground-truth citation recall; instead, we verify that the generated search terms, such as "transductive bandits" or "Fast Fourier Convolution inpainting", are a direct extraction of technical entities explicitly present in the user-provided excerpt and context.

\begin{figure*}[h!]
\begin{boxJ}
system: ...
\\ 

user:  You are now given an excerpt. Find me the paper cited in the excerpt, using the tools described above. Please make sure that the paper you select really corresponds to the excerpt: there will be details mentioned in the excerpt that should appear in the paper. If you read an abstract and it seems like it could be the paper we're looking for, read the paper to make sure. Also: sometimes you'll read a paper that cites the paper we're looking for. In such cases, please go to the references in order to find the full name of the paper we're looking for, and search for it, and then select it.

In the spirit of transductive bandits [CITATION] we consider a more general setting where answers are sets of arms. The set of actions and the set of answers can be different.
\\ 

assistant: \{"reason": "The excerpt mentions 'transductive bandits' as the key concept that the citation refers to. Since this is a specific term, I'll start by searching for papers on transductive bandits to find the foundational work.", "action": \{\\    "name": "search\_relevance",\\    "query": "transductive bandits"\\  \}\}
\\

user: ...
\\

assistant: \{"reason": "The excerpt introduces 'transductive bandits' as an existing concept and extends it to a setting where answers are sets of arms. Among the search results, Paper ID 4f0d485cbcde840533f23f0c8b0f3fa1ca2d74df directly introduces the transductive linear bandit problem, explicitly defining the scenario where measurement vectors (actions) and items (answers) can be different sets. This matches the excerpt's description of distinct action/answer sets. Other papers either apply transductive learning in different domains (e.g., text classification, video segmentation) or focus on variants like federated bandits without defining the core concept.",  "action": \{    "name": "select",    "paper\_id": "4f0d485cbcde840533f23f0c8b0f3fa1ca2d74df"  \}\}

\end{boxJ}
\caption{CiteGuard+DeepSeek Short trajectory (history length: 5)}
\label{citeguard: deepseek_short}
\vspace{-12pt}
\end{figure*}

\begin{figure*}[h!]
\begin{boxJ}
system: ...
\\ 

user:  You are now given an excerpt. Find me the paper cited in the excerpt, using the tools described above. Please make sure that the paper you select really corresponds to the excerpt: there will be details mentioned in the excerpt that should appear in the paper. If you read an abstract and it seems like it could be the paper we're looking for, read the paper to make sure. Also: sometimes you'll read a paper that cites the paper we're looking for. In such cases, please go to the references in order to find the full name of the paper we're looking for, and search for it, and then select it.

Our evaluation follows the protocol of [CITATION], a recent inpainting model that introduces a specialized architecture relying on Fast Fourier Convolutions [8]
\\ 

assistant: Okay, now I need to locate the paper that (a) is specifically an image-inpainting model and (b) centres its method on Fast Fourier Convolutions.  Because the excerpt does not specify the title, I'll start by issuing a highly-targeted search that includes both key phrases.\{"reason": "The excerpt identifies the missing citation as recent inpainting model that introduces a specialised architecture relying on Fast Fourier Convolutions [8].  The query must combine both aspects inpainting task and Fast Fourier Convolutions to find the exact paper.", "action": \{"name": "search\_relevance", "query": "Fast Fourier Convolution inpainting model"\}\}
\\

user: ...
\\

assistant: \{"reason": "The excerpt calls for the recent inpainting model whose key novelty is Fast Fourier Convolution, and abstract fdf7012ebe9d4c4d2d93004613e7a19f69a83a93 directly fits that description: it introduces a network that uses fast Fourier convolutions (FFCs) to obtain image-wide receptive fields for large-hole inpainting.",  "action": \{    "name": "select",    "paper\_id": "fdf7012ebe9d4c4d2d93004613e7a19f69a83a93"  \}\}
\end{boxJ}
\caption{CiteGuard+Kimi-K2 Short trajectory (history length: 5)}
\label{citeguard: kimi_short}
\vspace{-12pt}
\end{figure*}
\section{Difficulty Level Labels}
\label{app:difficulty_levels}
\begin{itemize}[itemsep=0pt]
    \item Easy: Several studies demonstrate the fragility of convolutional networks on simple corruptions. For example, [CITATION] apply impulse noise to break Google’s Cloud Vision API. (Ground-Truth: Google’s cloud vision api is not robust to noise)
   \item Medium: To address this, [CITATION] introduced Adversarial Filtering (AF). An overview is shown in Figure 2. The key idea is to produce a dataset D which is adversarial for any arbitrary split of (D\_train), (D\_test). (Ground-Truth: Swag: A Large-Scale Adversarial Dataset for Grounded Commonsense Inference)
   \item Medium-Hard: Even if we assume fixed filters using a combination of the above, our probabilistic formulation still allows learning the parameters of the GSM experts from data as outlined below. Consequently, we do not need to tune the trade-off weights between the brightness and gradient constancy terms by hand as in [CITATION]. (Ground-Truth: High Accuracy Optical Flow Estimation Based on a Theory for Warping)
   \item Hard: RCA [CITATION] is intermediate between PCA and LDA in its use of labeled data. Specifically, RCA makes use of so-called “chunklet” information, or subclass membership assignments. (Ground-Truth: Adjustment learning and relevant component analysis)
\end{itemize}

\section{Human Assessment On CiteGuard Alternative Citations}
\label{human_assessment_on_alternative_citations}

\begin{figure*}[h!]
\begin{boxJ}
Excerpt: For a comprehensive evaluation, we test on 19 datasets spanning a wide range of object, scene and fine-grained categories: ImageNet [18], StanfordCars [43], UCF101 [68], Caltech101 [25], Caltech256 [32], Flowers102 [56], OxfordPets [CITATION], Food101 [7], SUN397 [75], DTD [14], EuroSAT [37], FGVCAircraft [51], Country211 [61], CIFAR-10 [44], CIFAR-100 [44], Birdsnap [5], CUB [72], ImageNet-Sketch [73] and ImageNet-R [38]."\\

Oracle Reference: Cats and dogs (2012)

CiteGuard Suggestion: The truth about cats and dogs (2011)
\end{boxJ}
\caption{Example of CiteGuard Suggested Alternative Citations}
\label{alternative_citation2}
\vspace{-12pt}
\end{figure*}





Examples in Fig.~\ref{fig:citeagent_failure},~\ref{alternative_citation2}, and~\ref{alternative_citation}, suggest that CiteGuard’s extended retrieval actions and strategies not only improve the accuracy of original citation retrieval, but also expand the searching capacity to identify functionally equivalent references, supporting richer scholarly grounding with enhanced accuracy and robustness.
Importantly, our manual analysis (Table~\ref{tab:citeguard} and Figures~\ref{alternative_citation2} \&~\ref{alternative_citation}) reveals that CiteGuard is capable of both \textbf{lateral reasoning} (Fig.~\ref{alternative_citation2}) and \textbf{backward reasoning} (Fig.~\ref{alternative_citation}), behaviors that traditional citation retrieval systems typically lack.

\begin{itemize}[itemsep=0pt, parsep=0pt, topsep=5pt, partopsep=0pt]
    \item \textbf{Backward Reasoning:} While focusing on more recent publications, CiteGuard is capable of identifying citations of previous years written by the same author (Fig.~\ref{alternative_citation}).
    \item \textbf{Lateral Reasoning:} CiteGuard suggests peer or related work along with its identification of best-match citations (Fig.~\ref{alternative_citation2}), providing effective citation finding and alternative suggestions.
\end{itemize}

\section{LLM-As-A-Judge Failure}
\label{llm_as_a_judge_failures}
\paragraph{Evaluation Prompt.}

For the evaluation of OpenScholar citation attribution, we guide the LLM judge through the prompt in Fig. \ref{openscholar_prompt}.
\begin{figure}
\begin{boxJ}
As an Attribution Validator, your task is to verify whether a given reference can support the given claim. A claim can be either a plain sentence or a question followed by its answer. Specifically, your response should clearly indicate the relationship: Attributable, Contradictory or Extrapolatory. A contradictory error occurs when you can infer that the answer contradicts the fact presented in the context, while an extrapolatory error means that you cannot infer the correctness of the answer based on the information provided in the context. 
\ \\

Claim: {claim}
\ \\
Reference: {reference}
\end{boxJ}
\caption{OpenScholar citation attribution evaluation prompt to LLM}
\label{openscholar_prompt}
\vspace{-12pt}
\end{figure}

\paragraph{Failure Example} We show how an LLM judge can fail in its evaluation as a result of missing terminology nuances (Fig.\ref{example_llm_judge_failure}).

\begin{figure}[h!]
\begin{boxJ}
Claim: This technique of lower bounding mutual information is known as \textbf{Variational Information Maximization} [CITATION].
\ \\

Reference: The IM Algorithm: A \textbf{variational approach to Information Maximization}. …
\ \\

LLM judgement: Extrapolatory. Cannot infer the correctness of the answer based on the information provided in the context.
\end{boxJ}
\caption{LLM mistakenly judges a correct citation incorrect due to the slight difference in terminology}
\label{example_llm_judge_failure}
\vspace{-12pt}
\end{figure}

\section{LLM Generation Failure}

By examining LLM-generated outputs, we also observe failures due to their lack of important elements. For example, Fig. \ref{example_missing_citation} illustrates an LLM generation failure as a result of missing alternative citations.

\begin{figure}[h!]
\begin{boxJ}
However, LLMs … may lack \textbf{comparative analysis}, organizational structure, and key elements (Li et al. 2024). To address these challenges, …, \textbf{ChatCite}, … (Li et al. 2024).
\end{boxJ}
\caption{Example of an issue in a LLM-generated text: missing alternative citations (multiple papers other than ChatCite also address comparative analysis)}
\label{example_missing_citation}
\vspace{-12pt}
\end{figure}

\section{Examples of CiteGuard}
\subsection{Suggestion On Alternatives.}
\label{app:alternative_citation}
CiteGuard is capable of suggesting meaningful alternatives (Fig.\ref{alternative_citation2}, and Fig.\ref{alternative_citation}). An example of a case where alternative citations are not appropriate is as follows.
\begin{itemize}
    \item Zephyr-7B-Beta [CITATION] is an instruction-tuned version of Mistral-7B. (Ground-Truth: Zephyr: Direct Distillation of LM Alignment)
\end{itemize}

\subsection{Retrieval vs Long-Context.}
\label{app:long_context}
We present an example of the CiteGuard+GPT-4o agent when using the "read" action instead of the "find\_in\_text" action in Fig. \ref{ablation_long_context}, where the number of tokens can be as large as 4x. This is due to the additional tokens required when reading multiple full papers in context.

\begin{figure}
\begin{boxJ}
Excerpt: A second consideration is how to treat the image itself: the raw image could be fed directly into the reinforcement learning algorithm through a series of convolutions[CITATION].
\ \\

CiteGuard-Read:\\
Total Input Tokens(75900)\\
Select: Reward learning from human preferences and demonstrations in Atari
\ \\

CiteGuard-Find-In-Text:\\
Total Input Tokens(19182)\\
Select: Fully Convolutional Network with Multi-Step Reinforcement Learning for Image Processing
\end{boxJ}
\caption{CiteGuard example when using "read" vs "find\_in\_text"}
\label{ablation_long_context}
\vspace{-12pt}
\end{figure}

\begin{figure}
\begin{boxJ}
Excerpt: We learn $\pi$ using behavioral cloning [CITATION], which optimizes $\pi$ by minimizing the negative log-likelihood of actions given the images and language instructions.
\ \\

Oracle reference: \\
Alvinn: An autonomous land vehicle in a neural network (1988)
\ \\
CiteGuard suggestion: \\
A Framework for Behavioural Cloning (2001)
\end{boxJ}
\caption{Alternative citation suggested by CiteGuard, both are relevant according to human annotations.}
\label{alternative_citation}
\vspace{-12pt}
\end{figure}

\section{Human Annotators}
All human annotators are graduate students pursuing master’s or doctoral computer science degrees at universities where English is the primary language of instruction.

The annotators evaluated suggested citations along two axes:

\begin{itemize}[itemsep=0pt, parsep=0pt, topsep=5pt, partopsep=0pt]
    \item \textbf{Relevance:} Whether the cited paper genuinely supports the claim.
    \item \textbf{Sufficiency:} Whether the suggested citation can reasonably replace the original in scholarly writing.
\end{itemize}

Each human annotator was informed that the data collected would be used for a paper submission. The instruction given to the human annotators is as follows:
\\

Please review each excerpt below and:
\begin{enumerate}
    \item Select all papers that would be suitable for use as citations in the given excerpt context
    \item If none of the papers are suitable, please choose "None of the above"
\end{enumerate}

\section{Use of AI Assistants}
The AI assistant (i.e., Grammarly) is used for the writing of this manuscript. All content was critically reviewed and revised by human authors to ensure scientific accuracy and originality.